\documentclass[aps,pra,twocolumn,superscriptaddress,nofootinbib,longbibliography,floatfix]{revtex4-2}

\usepackage{graphicx}
\usepackage{amsmath}
\usepackage{multirow}
\usepackage{array}

\newcommand{\paneltagsize}{14.5}
\newcommand{\paneltag}[1]{{\ifdefined\pdfliteral\pdfliteral direct{2 Tr 0.28 w}\fi\textbf{\fontsize{\paneltagsize}{\paneltagsize}\selectfont #1}\ifdefined\pdfliteral\pdfliteral direct{0 Tr}\fi}}

\begin{document}

\title{Spectroscopy using dipole-phonon quantum logic}

\author{Lu Qi}
\thanks{These authors contributed equally.}
\affiliation{Duke Quantum Center, Duke University, Durham, NC 27701, USA}
\affiliation{Department of Electrical and Computer Engineering, Duke University, Durham, NC 27708, USA}

\author{Evan C. Reed}
\thanks{These authors contributed equally.}
\altaffiliation[Current address: ]{IonQ Inc., 4505 Campus Dr, College Park, MD 20740, USA}
\affiliation{Duke Quantum Center, Duke University, Durham, NC 27701, USA}
\affiliation{Department of Electrical and Computer Engineering, Duke University, Durham, NC 27708, USA}

\author{Duanyang Wang}
\affiliation{Duke Quantum Center, Duke University, Durham, NC 27701, USA}
\affiliation{Department of Physics, Duke University, Durham, NC 27708, USA}

\author{Boyan Yu}
\altaffiliation[Current address: ]{Department of Physics and Astronomy, University of British Columbia, 6224 Agricultural Rd, Vancouver, B.C., V6T 1Z1, Canada}
\affiliation{Duke Quantum Center, Duke University, Durham, NC 27701, USA}
\affiliation{Department of Electrical and Computer Engineering, Duke University, Durham, NC 27708, USA}

\author{Kenneth R. Brown}
\email{kenneth.r.brown@duke.edu}
\affiliation{Duke Quantum Center, Duke University, Durham, NC 27701, USA}
\affiliation{Department of Electrical and Computer Engineering, Duke University, Durham, NC 27708, USA}
\affiliation{Department of Physics, Duke University, Durham, NC 27708, USA}
\affiliation{Department of Chemistry, Duke University, Durham, NC 27708, USA}

\begin{abstract}
Dipole-phonon quantum logic (DPQL) offers novel approaches for state preparation, measurement, and control of quantum information in molecular ion qubits. Here we report the first experimental demonstration of DPQL as a spectroscopic tool using the shared motional mode of a CaO$^+$-Ca$^+$ ion chain to measure the frequency splitting of the ground $\Omega$-doublet of CaO$^+$. By comparing the dipole-phonon signal across three trap-frequency ramp ranges, we determine the splitting to be $\omega_{\text{mol}}=2\pi\times454\pm1$~kHz and place a lower bound on the effective dipole-phonon coupling. The rate of observation is limited by the low thermal population in the molecular ground rotational state.
\end{abstract}

\maketitle

\section{Introduction}

The coherent control of atoms has become an essential resource for studies of quantum physics and engineering quantum technologies.
Diatomic molecules, with their additional vibrational and rotational degrees of freedom, provide a broader spectrum that can be utilized for various purposes, including precision measurements~\cite{acme2018improved}, studies of quantum chemistry~\cite{doi:10.1126/science.adl6570, karman2024ultracold}, quantum simulation~\cite{li2023tunable,chen2023field} and quantum information processing~\cite{doi:10.1126/science.adf8999, doi:10.1126/science.adf4272}.
Molecular ions have also been used in these applications~\cite{doi:10.1126/science.adg4084, doi:10.1126/science.aan4701,doi:10.1126/science.aba3628,staanum2010rotational,germann2014observation,kortunov2021proton,calvin2018rovibronic,lin2020quantum}.
In molecular ion research, quantum logic spectroscopy (QLS) is a powerful tool for state preparation and measurement of internal states of single molecular ions~\cite{Chou2017,Wolf2016,doi:10.1126/science.aaz9837,holzapfel2024quantum}.
Originally developed for use in trapped atomic ion clocks~\cite{Schmidt_QLS_2005}, QLS transfers the quantum information from a trapped molecular ion's internal state to its motional state, which is then read out via a co-trapped atomic ion. 
This approach circumvents the need for molecular transitions with high photon cycling rates, often required for laser cooling in neutral molecule platforms~\cite{shuman2010laser}.
It is compatible with various molecular transitions, including hyperfine~\cite{Chou2017,holzapfel2024quantum}, rotational~\cite{doi:10.1126/science.aba3628,doi:10.1126/science.ado1001}, rovibrational~\cite{wu2026infrared}, and rovibronic transitions~\cite{Wolf2016,doi:10.1126/science.aaz9837}.
Recently, a novel approach to QLS was proposed that uses the interaction of the permanent dipole of a polar molecular ion with the trap electric field.
In this scheme known as dipole-phonon quantum logic (DPQL), opposite-parity states, such as \mbox{$\Lambda$}- and \mbox{$\Omega$}-doublet states~\mbox{\cite{Campbell_Hudson_2020}},  are used to facilitate the interaction.
For a molecule whose doublet splitting is comparable to the ions' secular frequency, the internal states can exchange energy directly with a motional mode~\cite{mills2020dipole}.
This technique offers a new set of tools for molecular state preparation and measurement. It has also been proposed for generating atom-molecule entanglement and mediating interactions between molecular ions through virtual phonons in an ion chain~\cite{Campbell_Hudson_2020}.

\begin{figure*}[t]
    \centering
    \includegraphics[width=\linewidth]{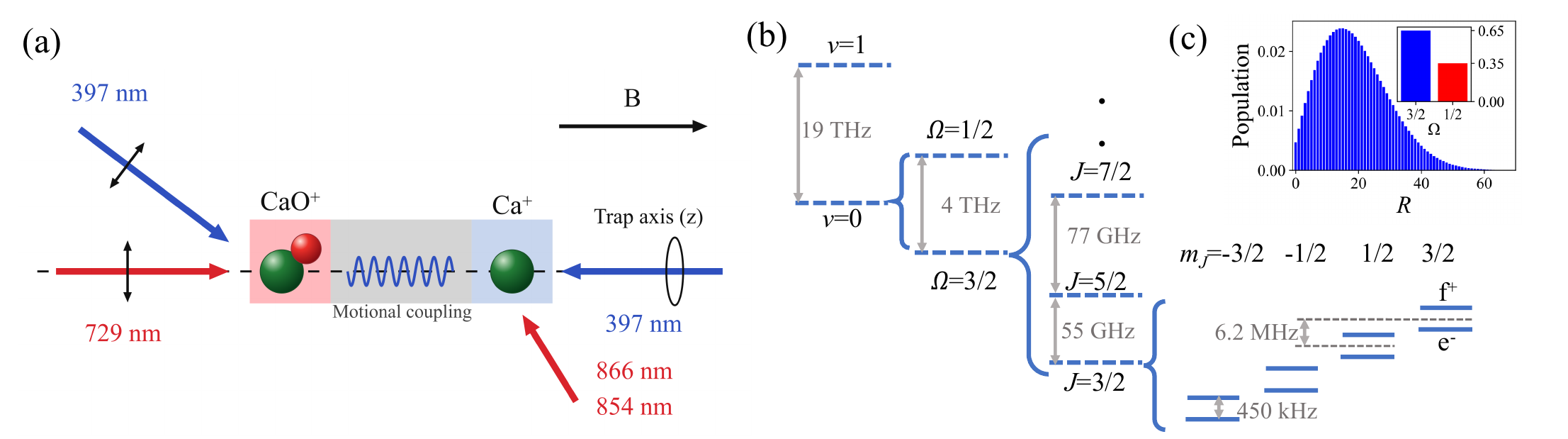}
    \caption{Experimental schematic and the energy structure and population distribution of CaO$^+$ at room temperature.
    (a) A schematic of the experimental setup for the demonstration of dipole-phonon quantum logic. 
    A single CaO$^+$ and Ca$^+$ are trapped along the trap axis. The schematic is not to scale; the ions are enlarged for clarity.
    A magnetic field of 6.2\,G is aligned to the trap axis ($z$) to provide the quantization axis and break state degeneracies. 
    The ion chain is shown in a top view, together with the laser wavelengths and orientations. The linearly polarized $397$\,nm beam and the $866$\,nm beam are used for Doppler cooling. EIT cooling uses both the linearly polarized and the $\sigma$-polarized $397$\,nm beam. The $729$\,nm laser drives the $S_{1/2}\leftrightarrow D_{5/2}$ operation, and the $854$\,nm laser quenches the $D_{5/2}$ population.
    (b) The energy level structure of CaO$^+$ ground electronic state X$^2\Pi$ (not to scale) in the presence of a 6.2\,G magnetic field. The ground state has a series of vibrational states. Each vibrational state is split by the fine structure. Each fine structure state couples with the molecular rotation to produce total angular momentum states.  These states are then split by the Zeeman sublevels which are finally split to $\Omega$-doublets. 
    e$^-$/f$^+$ indicates the odd/even parity state of the $\Omega$-doublets. 
    (c) Thermal population of the first 71 rotational states of the $\Omega=3/2$, $v=0$ manifold at room temperature, against the rotational quantum number $R=J-3/2$, so that $R=0$ is the DPQL-active $|\Omega=3/2,J=3/2\rangle$ level. Populations are normalized over the whole molecule. (Inset) The distribution of the two fine structure states, $\Omega=$1/2 and 3/2, at room temperature assuming Boltzmann distribution.
    }
    \label{fig:schematicofexperiment}
\end{figure*}

In this work, we present a spectroscopic measurement of the $\Omega$-doublet splitting of the calcium monoxide ion (CaO$^+$) using DPQL.
We previously reported evidence for a single dipole-phonon event in this system, with a lower bound on its statistical significance of $4.1\sigma$~\cite{Qi2024evidence}.
Here we analyze an independent and substantially larger dataset, together with both molecular and frequency-range controls, which raises the significance to $11.7\sigma$ and results in a measurement of the $\Omega$-doublet splitting.
In CaO\(^+\), the ground state is \(X^2\Pi_{3/2}\) and there is a low-lying electronically excited state \(A^2\Sigma^+\).
The combination of the rotational electronic Coriolis effect and the spin-orbit coupling leads to lifting of the degeneracy of the \(\Omega\) doublets, and the magnitude of this interaction increases with rotational number $R$ \cite{mills2020dipole}. 
Specifically, for the ground rotational state with total angular momentum $J=3/2$, 
the energy splitting between the lower energy doublet state $|e\rangle$ and higher energy doublet state $|f\rangle$ is calculated to be \(\omega_{\text{mol}} = 2\pi \times 450\pm35\,\text{kHz}\)~\cite{mills2020dipole} based on previously reported spectroscopic data of the electronic energy levels~\cite{VANGUNDY201817}.

In a linear Paul trap, CaO\(^+\) exhibits secular motion around its equilibrium position in the laboratory frame. In the rest frame of the ion, it experiences an oscillating electric field. 
When this oscillation frequency $\omega_{q}$ is near resonance with \(\omega_{\text{mol}}\), the interaction between the electric field and the molecular dipole can be described by the Jaynes-Cummings Hamiltonian in the rotating wave approximation~\cite{Campbell_Hudson_2020}:
\begin{equation}
     H_{\text{I}}/\hbar = \frac{\omega_{\text{mol}}}{2} \sigma_z + \omega_q \left(a_q^{\dagger} a_q + \frac{1}{2}\right) + \frac{g_q}{2} \left(a_q \sigma_{+} + a_q^{\dagger} \sigma_{-}\right),
    \label{eq:sysH}
\end{equation}
where \(g_q\) is the vacuum Rabi frequency. 

For the ground rotational state $|X^2\Pi_{3/2}, v=0, J=3/2, m_J\rangle$, \(g_q\) is calculated to be \(2\pi \times 2.0\,\text{kHz}\) for $|m_J|=3/2$ and  \(2\pi \times 0.67\,\text{kHz}\) for $|m_J|=1/2$ with the calculated permanent dipole moment $d_{\text{PDM}}=8.7$~Debye~\cite{mills2020dipole} and $\omega_{q}=\omega_{\text{mol}}=2\pi\times450$~kHz~(Appendix~\ref{app:fidelity}).
Prior to this work, neither the frequency of the \(\Omega\)-doublet nor the dipole-phonon coupling strength has been measured.
To cover the uncertainty in these parameters, we sweep the trap frequency at a constant rate through the predicted resonance.
By adiabatically ramping the trap frequency $\omega_q$ through resonance with \(\omega_{\text{mol}}\), the \(\Omega\)-doublet states will flip causing a corresponding change in the phonon state of the ion’s motion: $|f,n\rangle \leftrightarrow |e,n+1\rangle$, where $n$ is the phonon number of the Fock state. 
A calcium ion (Ca\(^+\)) is co-trapped with the CaO\(^+\) to provide sympathetic ground state cooling of the motion and readout of the phonon state, similar to the role of Ca\(^+\) in the QLS of CaH$^+$\cite{Chou2017}. 

\section{Dipole-phonon quantum logic}
A schematic of our experimental system is shown in Fig.~\ref{fig:schematicofexperiment}. In brief, a CaO$^+$-Ca$^+$ ion chain is prepared in a segmented blade trap in an ultra-high vacuum (UHV) chamber with pressure $\sim5\times10^{-11}$~Torr. (Appendix~\ref{app:cooling}) 
The two-ion chain has two modes of motion labeled in-phase (IP) and out-of-phase (OP) along each of the three axes.
The secular frequency of the axial IP mode is measured to confirm the identity of the CaO$^+$, by exciting the ions' motion with a resonant, oscillating electric field applied to one of the trap electrodes and observing the fluorescence change~\cite{10.1116/1.570570}.
The experiment begins by cooling both the radial and axial motional modes of the \(\mathrm{Ca}^+{-}\mathrm{CaO}^+\) ion chain. We use Doppler cooling followed by electromagnetically-induced transparency (EIT) cooling with phonon mode mixing (Appendix~\ref{app:cooling}) to prepare all six motional modes to have a mean phonon number of \(\bar{n} \approx 0.5\). 
A subsequent sideband cooling stage brings the axial IP and OP modes to \(\bar{n} \approx 0.02\).

\begin{figure*}[t]
    \centering
    \includegraphics[width=0.72\linewidth]{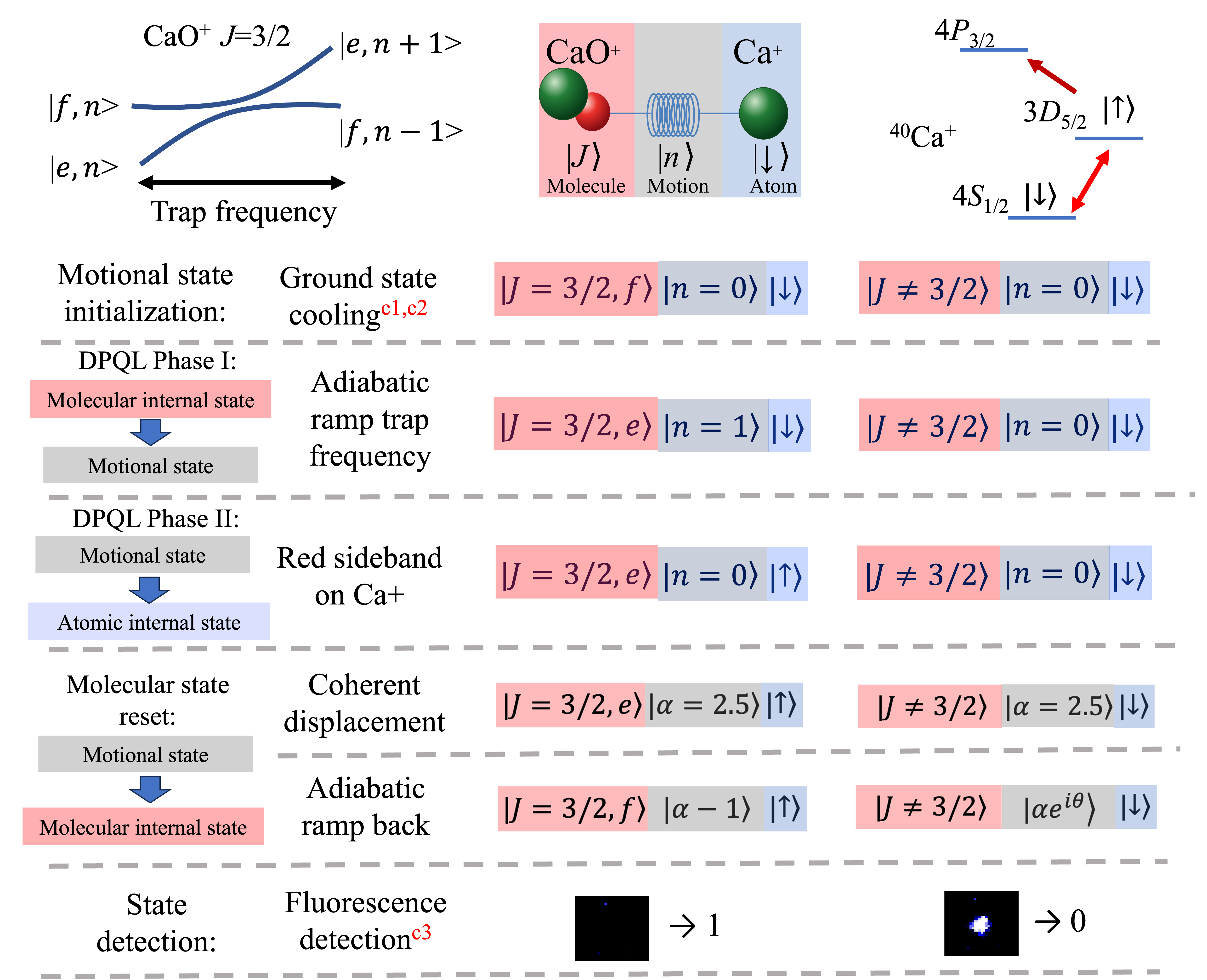}
    \caption{Experimental sequence for DPQL. The avoided crossing on the top left shows the state trajectories in the adiabatic passage operation. The energy structure of Ca$^+$ on the top right illustrates the atomic qubit states used in the experiment. 
    Shown here is every step of the experiment as well as the corresponding state changes of the molecular ion (both $J=3/2$ and $J\neq 3/2$), motion, and atomic ion. The results of single shot measurements of the Ca$^+$ ion are assigned the values of 1 for dark and 0 for bright. c1:\:Dark ion mass check. c2:\:Prior selection check of Ca$^+$ fluorescence. c3:\:Posterior selection check operation. See text for details.
    }
    \label{fig:DPQLsequence}
\end{figure*}

After the cooling stage, adiabatic ramping of the OP mode frequency from $\omega_{q,\text{high}}$ to $\omega_{q,\text{low}}$ is performed with a constant ramping speed over a duration of 8\,ms to induce the interaction between the $\Omega$-doublet opposite-parity states within the same $m_J$ manifold. 
When the ramping is over, a red sideband pulse on $S_{1/2} \leftrightarrow D_{5/2}$ of the Ca$^+$ ion maps the phonon state to the atomic state. 
If the interaction occurred, then a phonon would have been emitted into the motional mode, and the red sideband on the Ca$^+$ would successfully shelve the atom to the $D_{5/2}$ state. 
If there was no interaction, then the atom would remain in the $S_{1/2}$ state after the sideband operation.
Before detecting the atomic state, a calibrated RF pulse with frequency $\omega_{q,\text{low}}$ is applied to a DC electrode to displace the axial OP mode to a coherent state $|\alpha\rangle$~\cite{Meekhof_Wineland_1996} with $\alpha = 2.5$ (Appendix~\ref{app:fidelity}). 
The OP mode frequency is then adiabatically ramped back to $\omega_{q,\text{high}}$ with the same speed to reset the molecular ion’s internal state.
A coherent state is used here rather than the Fock state $|1\rangle$. The reset needs only a sufficient phonon number, not a particular number state. Preparing $|1\rangle$ would first require cooling the mode back to its ground state, which would lengthen every experiment. The extra motional excitation is harmless, as the red sideband pulse has already mapped the quantum information onto the Ca$^+$ spin.
At the end of the procedure, the atomic state is measured by driving the $S_{1/2} \rightarrow P_{1/2}$ transition and counting the scattered photons. 
The procedure for each experiment is shown in Fig.~\ref{fig:DPQLsequence}, and each experiment takes about 30~ms.
In an ideal experiment, a dark result indicates that the atomic ion is in $D_{5/2}$ and the occurrence of the dipole-phonon interaction, whereas a bright result indicates that the atomic ion is in $S_{1/2}$ state and that no interaction took place during the adiabatic ramping.

To determine $\omega_{\text{mol}}$ experimentally, we scan three overlapping OP frequency ranges [$\omega_{q,\text{high}}$, $\omega_{q,\text{low}}$] with the same ramping duration:  2$\pi\times$[534,454]~kHz (high), 2$\pi\times$[492,412]~kHz (middle), 2$\pi\times$[454,370]~kHz (low). 
Cooling and displacement steps are adjusted to the shifted secular frequencies while all other parts of the protocol are kept the same.
A dipole-phonon energy exchange requires the trap frequency to ramp fully through the doublet resonance, crossing it with margin on both sides, so only a range that contains the resonance in its interior produces a signal.
The pattern of signal and null across the three overlapping ranges therefore locates $\omega_{\text{mol}}$.

We apply three checks to reject false positives caused by background gas collisions and motional cooling failures.
The first check, applied every 10,000 experiments, is the mass check of the dark ion to confirm that the CaO$^+$ did not react with background gas to form a different molecule. 
The second check is applied at the beginning of the Doppler cooling stage to confirm the Ca$^+$ is fluorescing at the expected rate. 
If not, the experiment will halt until the check is passed. 
The third check is a posterior selection check and is applied after atomic state detection. 
This check includes a 854\,nm quenching pulse and a second atomic state detection. 
If a dark result is obtained in the experiment and a bright result is obtained in this check, the dark result is validated.
If the result of the check is also dark, the corresponding dark experiment result will be discarded.
Only data that pass all three checks are analyzed.
After these checks, the probability of a false positive result, a dark measurement without a dipole-phonon interaction, is 0.02. 
Conversely, when the molecule starts in the $\Omega$-doublet of the ground rotational state but does not result in a dark measurement, this is a false negative with probability $P_{\text{fn}}$. 
The probability of a false negative is a combination of imperfections mainly from motional decoherence in three operations: the fidelity of ramping with ground state $F_1(|m_J|)$, the fidelity of ramping with the coherent state $F_2(|m_J|)$, and the shelving and readout fidelity $P_d$ = 0.98. 
A single experiment starts with the molecule in the ground rotational state and in $|f\rangle$, with the mode in its motional ground state. The first ramp and the readout give a dark result with probability $F_1(|m_J|)\,P_d$. The return ramp carries the molecule from $|e\rangle$ back to $|f\rangle$ with fidelity $F_2(|m_J|)$, preparing the next cycle. Iterating to steady state at the calculated values $d_{\text{PDM}}=8.7$~Debye and \(\omega_{\text{mol}} = 2\pi \times 450\,\text{kHz}\) gives $p_s=0.71$ for the $|m_J|=3/2$ manifold and $0.21$ for $|m_J|=1/2$, the latter smaller because of its weaker coupling to the phonon mode (Appendix~\ref{app:fidelity}).

For CaO\(^+\), the probability of population of the ground rotational state at room temperature is only 0.47\% due to its small rotational constants as shown in Fig.~\ref{fig:schematicofexperiment}(c).
Preparing the molecular qubit by projective measurement therefore requires many repetitions~\cite{Vogelius_2006}.
Upon successful projective measurement, the ground state population will rethermalize with a predicted $1/e$ time of 4 seconds from our rate-equation model (Appendix~\ref{app:rotational}).
We therefore expect to measure multiple successive instances of the dipole-phonon interaction before rethermalization.
We refer to these bursts as DPQL signals.
The chemical lifetime of \(\mathrm{CaO}^+\) in our system is limited primarily by the reaction between CaO$^+$ and background gases that produce \(\mathrm{CaOH}^+\), which does not have the correct internal structure to produce a dipole-phonon interaction. Across the data collected in this work, the measured lifetimes of CaO$^+$ are consistent with an exponential distribution with a characteristic chemical lifetime of $74\pm10$~min.
During the \(\mathrm{CaO}^+\) lifetime, we repeat the experiment hundreds of thousands of times and pause every 60{,}000 times (30~min) to recalibrate the laser frequency to compensate for 729~nm laser frequency drift.

\section{Control experiment}

To determine whether the bursts of frequent shelving events observed with the CaO$^+$-Ca$^+$ ion chain are DPQL signals, we performed a control experiment by applying the DPQL search procedure to CaOH$^+$ after the CaO$^+$ reacts to form CaOH$^+$.
Because CaOH$^+$ has a closed-shell $^1\Sigma$ ground state, it does not support an $\Omega$-doublet and therefore cannot exhibit the internal structure required for DPQL. 
Moreover, the mass difference between CaOH$^+$ and CaO$^+$ produces only a small shift of the secular frequencies, $1.3$~kHz for the IP mode and $1.2$~kHz for the OP mode at $\omega_{q,\text{high}}=2\pi\times492$~kHz. We calibrate these shifts in the experiment, and they do not affect the quality of the EIT cooling, the ground-state cooling, or the adiabatic ramping sequence. 

\begin{figure*}[t]
    \centering
    \includegraphics[width=\linewidth]{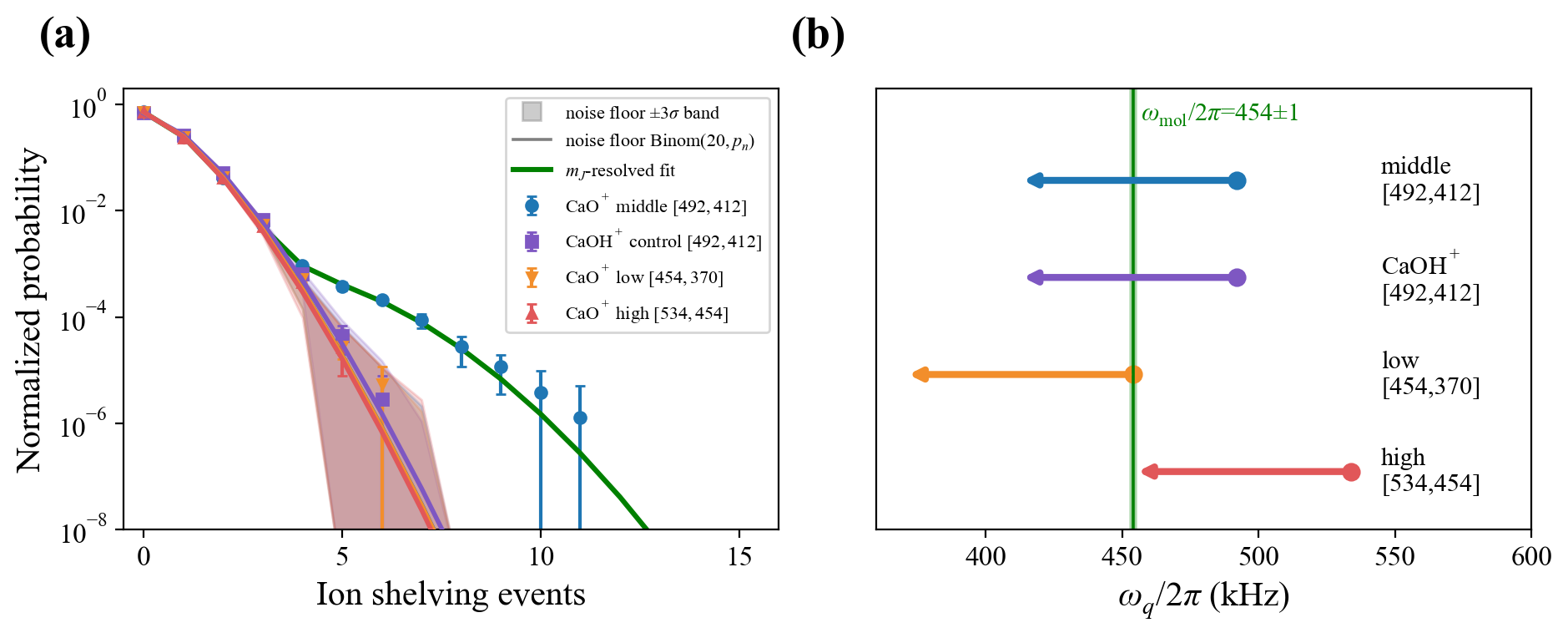}
    \caption{Dipole-phonon signal and the frequency-scan geometry.
    (a) Normalized probability of observing a given number of ion-shelving events in 20-experiment windows, for all four datasets: the CaO$^+$ middle range (blue circles), the CaOH$^+$ control molecule over the same range (purple squares), and the CaO$^+$ low (orange down triangles) and high (red up triangles) ranges. Each is shown with its own noise floor (gray line, binomial at that dataset's false-positive rate $p_n$) and the $\pm3\sigma$ spread obtained from scrambled detection records (gray band). The green curve is the $m_J$-resolved fit to the CaO$^+$ middle-range data, with a total ground-rotational-state population of $0.60^{+0.14}_{-0.11}\%$ and a per-event shelving probability of $0.17$ for the $|m_J|=3/2$ manifold.
    (b) Frequency geometry of the four datasets: each arrow spans the (downward) trap-frequency ramp of the CaO$^+$ middle range, the CaOH$^+$ control molecule over the same range, and the CaO$^+$ high and low ranges. The start of each arrow ($\bullet$) marks where the motional mode is sideband-cooled and its temperature characterized. The green band is the doublet frequency isolated by the joint analysis of Fig.~\ref{fig:rangebound}, $\omega_{\text{mol}}=2\pi\times454\pm1$~kHz.}
    \label{fig:newcompare}
\end{figure*}
 
\begin{figure*}[t]
    \centering
    \includegraphics[width=\linewidth]{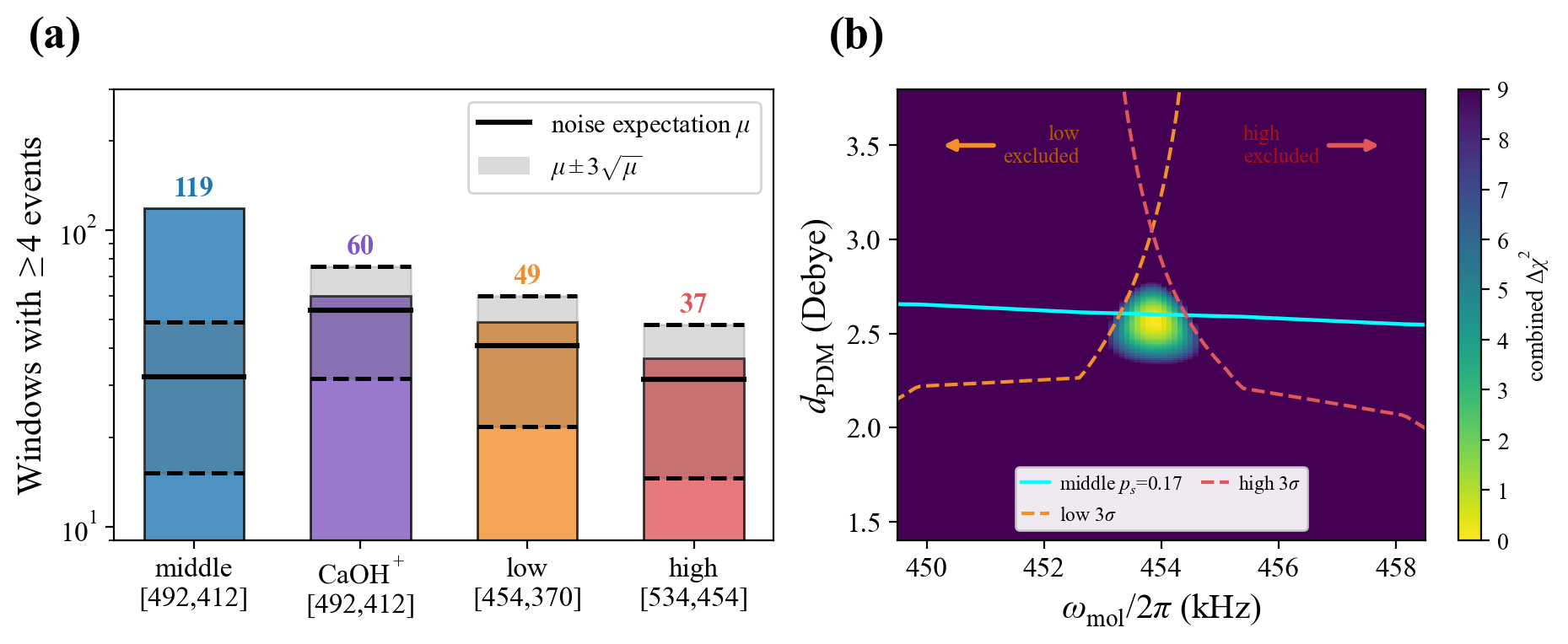}
    \caption{Statistical significance and the joint frequency--coupling constraint.
    (a) Number of 20-experiment windows containing $\geq4$ ion-shelving events for each dataset. Experimental values are denoted by the height of the colored bars and the number. The solid black line is the average count, $\mu$, expected from that dataset's noise floor and the shaded band is $\mu\pm3\sqrt{\mu}$. The noise floor is evaluated run by run and then combined over all runs of the dataset to calculate $\mu$ (Appendix~\ref{app:tail}). 
    (b) Joint constraint on the effective dipole-phonon coupling $d_{\text{PDM}}$ and the doublet frequency $\omega_{\text{mol}}$. The color is the combined $\Delta\chi^2$ of Appendix~\ref{app:joint}, which compares the simulated per-event signal of each range with the data. The cyan curve marks where the middle range reproduces the measured shelving probability $p_s=0.17$, and the dashed curves mark the tail-based detection limits of the low and high ranges. No confidence contour is drawn, since the coverage of this score is not calibrated. The low and high ranges exclude the corresponding sides (arrows), isolating $\omega_{\text{mol}}\approx2\pi\times454$~kHz and $d_{\text{PDM}}\approx2.56$~Debye.}
    \label{fig:rangebound}
\end{figure*}

\section{Results and discussion}

The middle-range data show an excess of shelving events that is absent from the high- and low-range data and the CaOH$^+$ control. Comparing the three ramp ranges gives $\omega_{\text{mol}}=2\pi\times454\pm1$~kHz.
In Fig.\,\ref{fig:newcompare}, we plot the shelving-event statistics of all four datasets.
The datasets contain 157 middle-range runs, 165 control-molecule runs, 162 low-range runs, and 139 high-range runs.
Each run contains 10{,}000 time-ordered experiments.
The dipole-phonon interaction increases the probability of shelving Ca$^+$ in the $D$ state.
We therefore search for the interaction by counting shelving events in windows of 20 experiments.
Figure~\ref{fig:newcompare}a shows the normalized distribution of shelving counts in 20-experiment windows.
Due to the low probability that the ground rotational state is populated at room temperature and its short lifetime, we use the average shelving probability of the total dataset as the noise floor.
This is elevated from zero due to systematic noise and operational errors that produce false positives in the search procedure.
To estimate the standard deviation of the noise floor, we randomly permute the order of each dataset one thousand times, which removes temporal correlations while keeping the false positive rate fixed.
The resulting $\pm3\sigma$ band is plotted around the corresponding noise floor of each dataset.
The middle-range distribution exceeds the noise floor by more than three standard deviations at large shelving counts.
The control-molecule, high-range, and low-range datasets show small deviations from the noise predictions at low shelving counts ($k\leq3$), but none shows a high-count excess comparable to that in the middle-range data.
For $k\geq6$, the middle-range data contain 30 windows, compared with a noise expectation of 0.07; the control-molecule, high-range, and low-range datasets contain 0, 0, and 1 window, respectively.

We fit the shelving-event distribution of Fig.~\ref{fig:newcompare}a over non-overlapping $20$-shot windows by a Poisson likelihood, modelling it as a mixture of an inactive, noise-only population and a DPQL-active one. The fit gives a steady-state per-event shelving probability $p_s=0.17\pm0.01$ for an active population of $0.3\%$, with a fitted noise floor $p_n=0.017$. We identify this population with the $|m_J|=3/2$ sublevels of the $J=3/2$ ground rotational state. The angular-momentum relation fixes the $|m_J|=1/2$ coupling to be one third of the $|m_J|=3/2$ value.
As a result, the $|m_J|=1/2$ shelving probability is predicted to be $0.02$, and indistinguishable from the noise floor (Appendix~\ref{app:tail}). Assuming the two manifolds are equally populated, the total ground-rotational-state population is $0.60^{+0.14}_{-0.11}\%$, consistent with the independent Boltzmann estimate of $0.47\%$ for the $J=3/2$ ground rotational state.
The measured shelving probability lies well below the value of $0.71$ predicted for the $|m_J|=3/2$ manifold at the calculated dipole moment. One possible explanation is a reduced dipole-phonon coupling. Interpreting the entire reduction this way through the steady-state model gives $d_{\text{PDM}}=2.59\pm0.11$~Debye, compared with the calculated $8.7$~Debye.

We quantify the significance of this signal with a tail-counting test. For each dataset we count the number of $20$-experiment windows containing $\geq4$ ion-shelving events, a tail that the noise floor produces only rarely (Fig.~\ref{fig:rangebound}a). Each dataset's own noise floor sets the number of such windows expected in the absence of signal, $\mu$. Each count is compared with the band $\mu\pm3\sqrt{\mu}$ (Fig.~\ref{fig:rangebound}a). 
With a Poisson background model, observing at least 119 tail windows in the middle-range data gives $p\approx6\times10^{-32}$, or a nominal one-sided significance of $11.7\sigma$. The control-molecule, low-range, and high-range data each lie within $1.2\sigma$ of their expectations. The noise floor is evaluated run by run to avoid underestimating the expected tail with a single averaged noise floor (Appendix~\ref{app:tail}).

Combining the fitted per-event shelving probability with the signal/null pattern of the three ramp ranges yields a joint constraint on the doublet splitting and the effective dipole-phonon coupling (Fig.~\ref{fig:rangebound}b). The joint fit gives $\omega_{\text{mol}}=2\pi\times454\pm1$~kHz and, with the same attribution of the shortfall in $p_s$, an effective coupling of $d_{\text{PDM}}=2.56\pm0.07$~Debye, consistent with the value inferred above from the shelving-event distribution.
Our steady-state model incorporates the ramp geometry, sideband pumping, and motional dephasing (Appendix~\ref{app:fidelity}).
However, imperfections outside the model also lower the measured shelving probability at fixed dipole. These include drifts in the trap RF potential and in the axial micromotion compensation, which keeps the ion at an electric-field null as the axial potential is ramped. Any of these would account for part of the shortfall in $p_s$ and imply a larger true dipole, so the inferred value of $2.56$~Debye is an effective lower bound rather than a direct measurement of the permanent dipole moment.

\begin{figure*}[t]
    \centering
    \includegraphics[width=\linewidth]{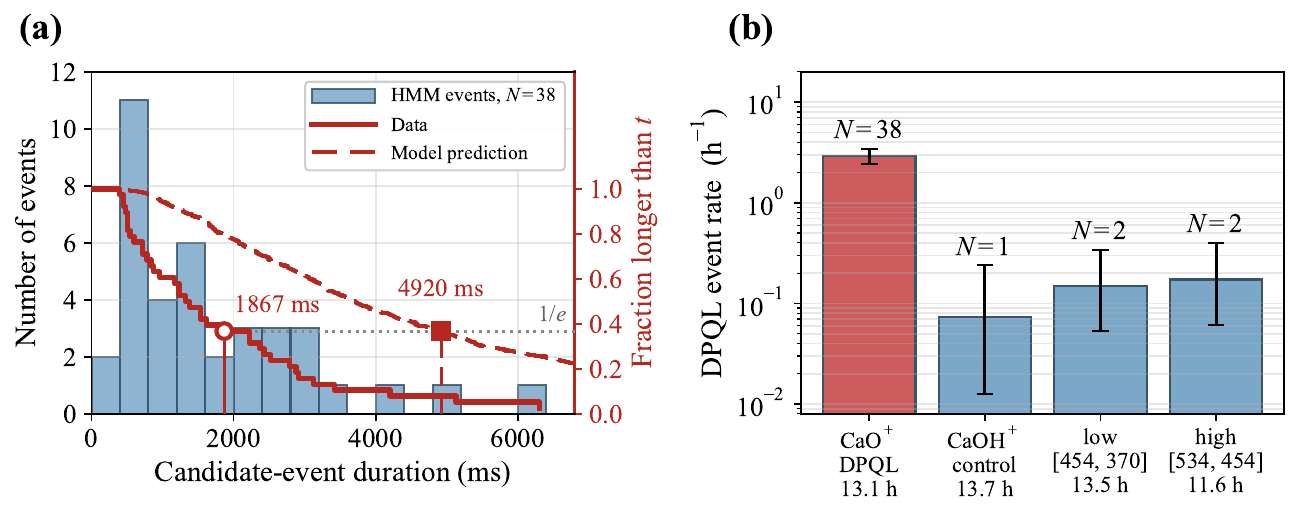}
    \caption{Signals identified by the parameter-free HMM.
    (a) Durations of the $38$ candidate events found in the CaO$^+$ DPQL search data. Left axis: histogram, bin width $400$~ms. Right axis: the survival function of the data (solid) and of the model prediction (dashed), the latter obtained by decoding the Monte-Carlo simulation with the same HMM and the same dual-threshold criterion. Markers give the $1/e$ crossing of each, $1867$~ms and $4920$~ms.
    (b) Event rate of each dataset under the same dual-threshold criterion (boundaries at posterior $0.5$, selection at peak posterior $\geq0.9$), with $68\%$ Poisson intervals. The two off-resonant ranges supply the background reference.}
    \label{fig:hmmsummary}
\end{figure*}

To identify individual bursts, we use a hidden Markov model (HMM, Appendix~\ref{app:HMMnew}) that was built directly from the blackbody-driven rotational--vibrational rate matrix (Appendix~\ref{app:rotational}).
The HMM assigns each measurement a posterior probability of DPQL activity, accounting for false-positive and false-negative detection outcomes.
Signals are read out from the posterior probability and are selected by a dual-threshold (hysteresis) rule. A candidate is a contiguous stretch with $P(\Omega|x)>0.5$, which sets its boundaries, and it is retained if its peak posterior reaches $0.9$.
In total, the HMM identified 38 DPQL signals in the $13.07$~h of CaO$^+$ middle-range data, compared with 2 and 2 in the $25.06$~h of low- and high-range data and 1 in the $13.73$~h of control-molecule data (Fig.\,\ref{fig:hmmsummary}b).
These counts are lower bounds, because the dual-threshold criterion recovers $78\%$ of the signals in simulations of the model, in which the true state of the molecule is known (Appendix~\ref{app:HMMnew}).
We collect all HMM-identified signals and plot the distribution of their durations in Fig.\,\ref{fig:hmmsummary}a.

Their survival function falls to $1/e$ at $t_{1/e}=1867^{+503}_{-337}$~ms. To compare this with the model we pass the Monte-Carlo simulation of Appendix~\ref{app:rotational} to the same HMM under the same dual-threshold criterion. The durations it returns have $t_{1/e}=4920$~ms (Fig.\,\ref{fig:hmmsummary}a).
The measured durations are a factor of $2.6\pm0.6$ shorter than the model predicts, the model being driven by blackbody radiation at $297$~K (Appendix~\ref{app:rotational}). A smaller dipole moment would only widen this gap, since the blackbody rates scale as its square.
One possible explanation is that the electromagnetic radiation environment at the trap center can differ due to thermal heating of trap electrodes and the trap geometry changing the electromagnetic environment for wavelengths longer than the trap dimensions \cite{doi:10.1126/science.ado1001,7ypf-91jr}.
Rotational and rovibronic spectroscopy of molecular ions could be used to accurately probe the electromagnetic environment of the trap in the future \cite{doi:10.1126/science.ado1001,doi:10.1021/acs.jpca.5c00229}.

$
$

\section{Conclusion}
In conclusion, we use dipole-phonon quantum logic to perform spectroscopy of the ground rotational state \mbox{$\Omega$}-doublet of CaO\mbox{$^+$} in a room-temperature ion trap, probing its coupling to a shared phonon mode of the trapped-ion chain.
By sweeping the trap frequency through the molecular resonance, we observe the dipole-phonon interaction in the middle-range data at a statistical significance of $11.7\sigma$.
Neither the two off-resonant ramp ranges nor the CaOH$^+$ control reproduces the middle-range shelving excess.
The three-range comparison gives a ground-state $\Omega$-doublet splitting of $\omega_{\text{mol}}=2\pi\times454\pm1$~kHz, in agreement with the \emph{ab initio} prediction of Ref.~\cite{mills2020dipole}, and an effective dipole-phonon coupling bound of $d_{\text{PDM}}\geq2.5$~Debye. To our knowledge, this is
the first measurement of the CaO$^+$ ground-state $\Omega$-doublet splitting.
The rate and lifetime of the DPQL signals are smaller than expected. 
Further study is needed to understand this discrepancy which is likely due to the difference between local thermal radiation and ideal blackbody radiation model.
In future experiments, a cryogenic ion trap could increase the ground rotational state population and lifetime while reducing background collisions~\cite{7ypf-91jr}.
The resulting increase in signal rate would allow more frequent spectroscopic measurements of the ground-state $\Omega$-doublet.
Improved state preparation could also allow coherent control of molecular internal states, including higher rotational states with stronger dipole-phonon coupling, for quantum information and sensing.

\begin{acknowledgments}
This work is supported by a collaboration between the U.S. DOE and other Agencies. This material is based upon work supported by the U.S. Department of Energy, Office of Science, National Quantum Information Science Research Centers, Quantum Systems Accelerator. Additional support is acknowledged from the Army Research Office (W911NF-21-1-0346 and W911NF-26-1-A063) and the ARO Spectator Qubit MURI (W911NF-18-1-0218). We thank Hao Wu and Michael Heaven for providing theoretical calculations of energy structures of CaO$^+$. We thank Eric Hudson and Wes Campbell for helpful discussions.
\end{acknowledgments}

\appendix
\section{Methods}
\subsection{Three dimensional cooling of ion chain}\label{app:cooling}
To prepare the CaO\(^+\)-Ca\(^+\) ion chain, we first load two Ca\(^+\) ions into the segmented blade trap in an ultra-high vacuum (UHV) chamber~\cite{Qi2023}. 
We then introduce O\(_2\) gas via a leak valve while the 397\,nm laser is tuned closer to the \(S_{1/2} \rightarrow P_{1/2}\) resonance of Ca\(^+\) to increase the population in the \(P_{1/2}\) state to facilitate the reaction between Ca$^+$ and O$_2$~\cite{Schmid_2019,mills2020dipole}.
During this process, the vacuum pressure is maintained at \(1.5 \times 10^{-9}\,\text{Torr}\). 
The leak valve is closed once one of the two trapped Ca$^+$ ions reacts with O$_2$, thereby forming CaO$^+$. The pressure returns to about \(5 \times 10^{-11}\,\text{Torr}\)  within 5 minutes after the leak valve is closed.
The vacuum pressure is deduced from a combination of ion pump reading and the ion position exchange rate~\cite{OKA2021111482}.

To cool the ion-chain motion near the ground state ($\bar{n}<1$), we use a combination of EIT cooling and phonon-mode mixing. The mass difference between Ca$^+$ and CaO$^+$ makes the participation ratios of the two ions in the radial modes differ strongly. The radial modes are therefore nearly uncoupled and oscillate almost independently. At an axial OP mode frequency of $2\pi\times492$~kHz, the mass-weighted participation ratio of a radial mode is $99.5\%$ on one ion and $0.5\%$ on the other. 
To overcome this, we mix the radial out-of-phase (OP) mode with the axial OP mode\cite{PhysRevA.89.062332,PhysRevLett.119.193602} to enable efficient cooling of both. We apply a small DC bias in the radial plane (x–y). Due to the mass difference, the two ions are displaced by different amounts from the RF null, which tilts the ion chain away from the trap axis $z$ \cite{Qi2023}. This tilt introduces cross terms in the potential such as $x\cdot z$ and $y\cdot z$, which couple the axial and radial modes. When the axial OP frequency is tuned close to one of the radial OP frequencies, the interaction in the rotating-wave approximation can be written as
$
H_{\text{int}} = g(\theta_i)\,\bigl(a^\dagger b + a b^\dagger\bigr),
$
where $g(\theta_i)$ is the coupling strength, $\theta_{i=x/y}$  is the angle between the ion chain and the x/y axis, $a$ and $b$ are the annihilation operators of the two coupled modes. By setting mode frequency detuning $\Delta \approx 0$, we realize an effective 50:50 beam-splitter interaction. Cooling the mixed mode then cools both the axial and the radial components.

The cooling sequence is as follows. We first set the axial OP frequency close to the higher radial OP frequency (the two radial OP modes are separated by about $2\pi\times$40~kHz, at $2\pi\times$690~kHz and $2\pi\times$650~kHz, respectively) and apply Doppler cooling. We then use EIT cooling on the radial and the axial in-phase (IP) modes respectively to reach $\bar{n}\approx 0.5$ and suppress Debye–Waller effects. Next, we apply EIT cooling to the axial–radial mixed modes to bring them to $\bar{n}\approx 0.5$.
At the end of this procedure, all motional modes are cooled close to the ground state with $\bar{n}\approx 0.5$. We then ramp the axial OP mode to $2\pi\times$492~kHz ($2\pi\times$100~kHz/ms) and apply resolved-sideband cooling to both the axial IP and OP modes, reaching $\bar{n}\approx 0.02$ for both axial modes. This prepares the motion for the adiabatic ramps.

\subsection{Estimation of operational fidelity}\label{app:fidelity}

We estimate the per-event shelving probability by simulating the full DPQL cycle with a Lindblad master equation,
\begin{equation}
    \dot{\rho}=-\frac{i}{\hbar}[H_{\text{I}}(t), \rho]+\sum_k\left(L_k \rho L_k^{\dagger}-\frac{1}{2}\left(L_k^{\dagger} L_k \rho + \rho L_k^{\dagger} L_k\right)\right) \,  .
    \label{masterEq}
\end{equation}
$H_{\text{I}}(t)$ is the Jaynes--Cummings Hamiltonian of Eq.\,\ref{eq:sysH}. We include two dissipators: motional dephasing, $L_{\text{deph}}=\sqrt{\Gamma_q}\,a^{\dagger}a$ with rate $\Gamma_q=2/T$ and measured coherence time $T=4$~ms; and resolved-sideband cooling at rate $\Gamma_{\text{RSB}}=10$~ms$^{-1}$, represented by a jump operator that removes one phonon from the system. This dissipator models the ground-state cooling applied at the start of each experiment. When the cooling frequency lies near $\omega_{\text{mol}}$, it combines with the off-resonant Jaynes--Cummings coupling to pump the molecule toward the lower doublet state.
Axial OP-mode heating is negligible on the ramp timescale~\cite{Qi2023} and is omitted.

\begin{figure*}[t]
    \centering
    \includegraphics[width=\linewidth]{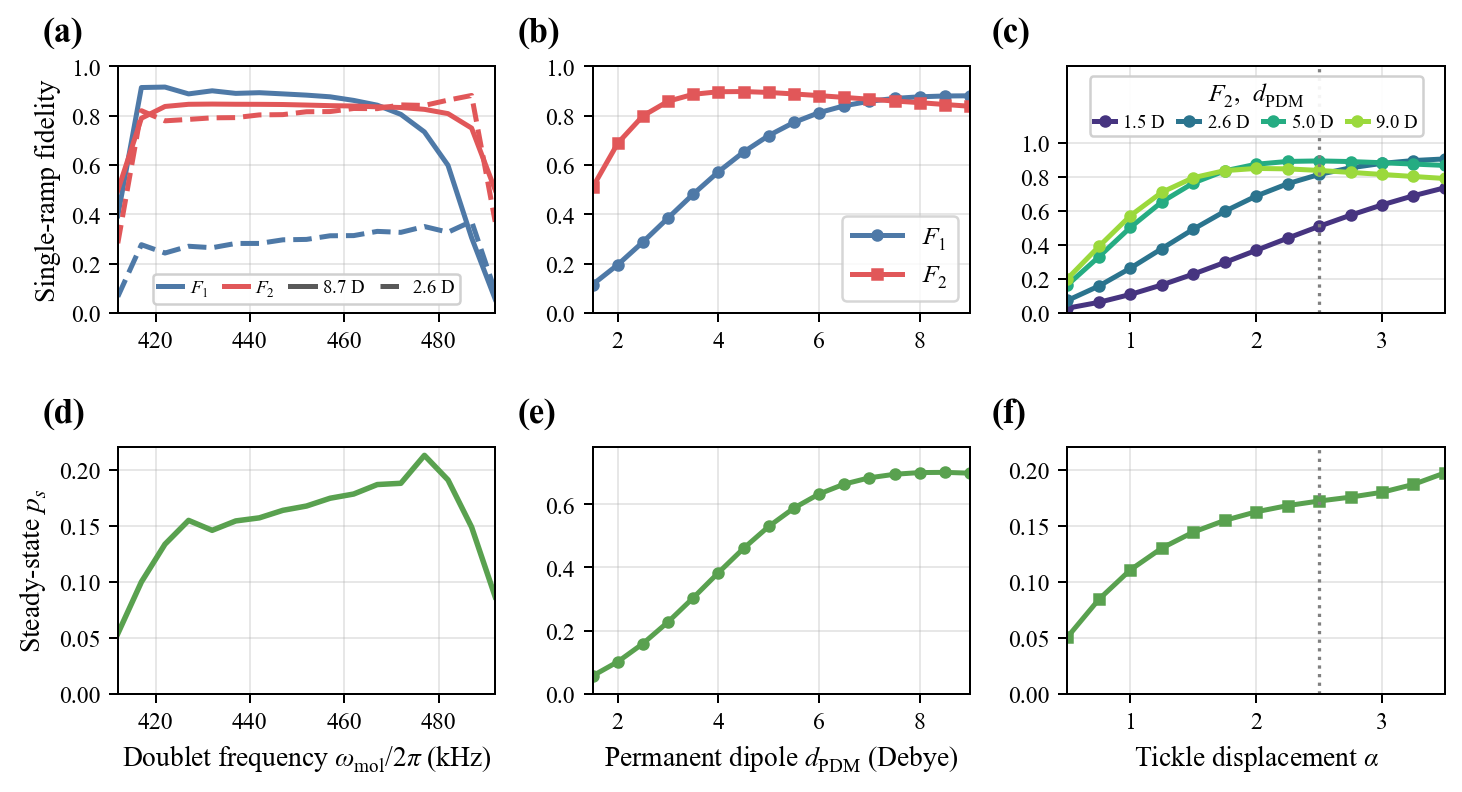}
    \caption{Simulated single-ramp fidelities (top) and steady-state per-event shelving probability $p_s$ (bottom) versus the model parameters, for the $|m_J|=3/2$ manifold and the middle-range ramp $2\pi\times[492,412]$~kHz at $2\pi\times10$~kHz/ms. (a) First- and second-ramp fidelities $F_1$ (blue) and $F_2$ (red) versus the doublet frequency $\omega_{\text{mol}}$, at the calculated permanent dipole $d_{\text{PDM}}=8.7$~Debye (solid) and a reduced $2.6$~Debye (dashed); both fidelities fall as the resonance approaches a ramp edge, and near the cooling frequency ($492$~kHz) $F_1$ is further suppressed by sideband pumping. (b) $F_1$ and $F_2$ versus $d_{\text{PDM}}$ at $\omega_{\text{mol}}=2\pi\times454$~kHz. (c) $F_2$ versus the tickle displacement $\alpha$ for several $d_{\text{PDM}}$.
    The displacement dependence varies with the coupling strength. The dotted line marks the experimental choice $\alpha\approx2.5$. (d)--(f) Steady-state $p_s$ versus $\omega_{\text{mol}}$ (at $d_{\text{PDM}}=2.6$~Debye), $d_{\text{PDM}}$ (at $\omega_{\text{mol}}=2\pi\times454$~kHz), and $\alpha$ (at $d_{\text{PDM}}=2.6$~Debye). Because each molecule is interrogated repeatedly, $p_s$ is set jointly by the two ramp fidelities and lies below the single-ramp values.}
    \label{fig:fidmech}
\end{figure*}

The vacuum Rabi frequency in $H_{\text{I}}(t)$ is $g_q=d_{\text{ef}}(J,m_J)\,E_q / \hbar$, where $d_{\text{ef}}(J,m_J)$ is the transition dipole between $\Omega-$doublets of the same $m_J$. For $|X^2\Pi_{3/2},v=0, J,m_J\rangle$ states of CaO$^+$~\cite{bohn2009electricdipolesultralowtemperatures},
\begin{equation}
    d_{\text{ef}}(J,m_J)\equiv \langle J,m_J,e/f|\hat{\mathbf d}|J,m_J,f/e\rangle
= \mu_{\text{pd}}\,\frac{|m_J|\,\Omega}{J(J+1)},
\end{equation}
with $\mu_{\text{pd}}=8.7$~Debye the calculated permanent dipole moment~\cite{mills2020dipole}, and $E_q=b_q \sqrt{2\hbar m_{\text{CaO}}\omega_{\text{mol}}^3/e^2}$ the single-phonon electric-field amplitude of mode $q$ (normal-mode eigenvector component $b_q$, CaO$^+$ mass $m_{\text{CaO}}$, elementary charge $e$). For the ground rotational state ($J=3/2$, $\omega_{\text{mol}}=2\pi\times450$~kHz), $g_q=2\pi\times2.0$~kHz for $|m_J|=3/2$ and $2\pi\times0.67$~kHz for $|m_J|=1/2$.
The dependence of the single-ramp fidelities $F_1$ and $F_2$ on the doublet frequency, permanent dipole, and displacement $\alpha$ is shown in Fig.~\ref{fig:fidmech}(a--c).

In the experiment each molecule is interrogated repeatedly, and the state carried from one cycle to the next reaches a steady state. We define $p_s$ as the per-event shelving probability in that steady state, and it is $p_s$ that is fit to the data in the main text. The second ramp must return the molecule to $|f\rangle$ for the following cycle, so the saturated $p_s$ is set jointly by the two ramp fidelities and by this repeated preparation, and is lower than its single-cycle value.
We plot how $p_s$ is affected by $\omega_{\text{mol}}$, $d_{\text{PDM}}$, and $\alpha$ in Fig.~\ref{fig:fidmech}(d--f).

\section{Rotational dynamics of CaO$^+$}\label{app:rotational}
We model the internal-state populations of CaO$^+$ with a rate-equation model driven by $297$\,K blackbody radiation (BBR), taking the equilibrium populations to follow the Boltzmann distribution $p_i\propto g_i e^{-E_i/k_BT}$. The model takes a vibrational constant $\omega_e=634$\,cm$^{-1}$, a spin-orbit constant of $130$\,cm$^{-1}$, which splits the $\Omega=1/2$ and $\Omega=3/2$ fine-structure manifolds, and a rotational constant $B_e=0.37$\,cm$^{-1}$. At $T=297$\,K these place $95\%$ of the population in the ground vibrational state and $65\%$ ($35\%$) in the $\Omega=3/2$ ($\Omega=1/2$) manifold, leaving $0.47\%$ in the DPQL-active ground rotational state $|\Omega=3/2,J=3/2\rangle$.

\begin{figure*}[t]
    \centering
    \begin{minipage}[b]{0.431\linewidth}
        \raggedright\paneltag{(a)}\\[0.4ex]
        \includegraphics[width=\linewidth]{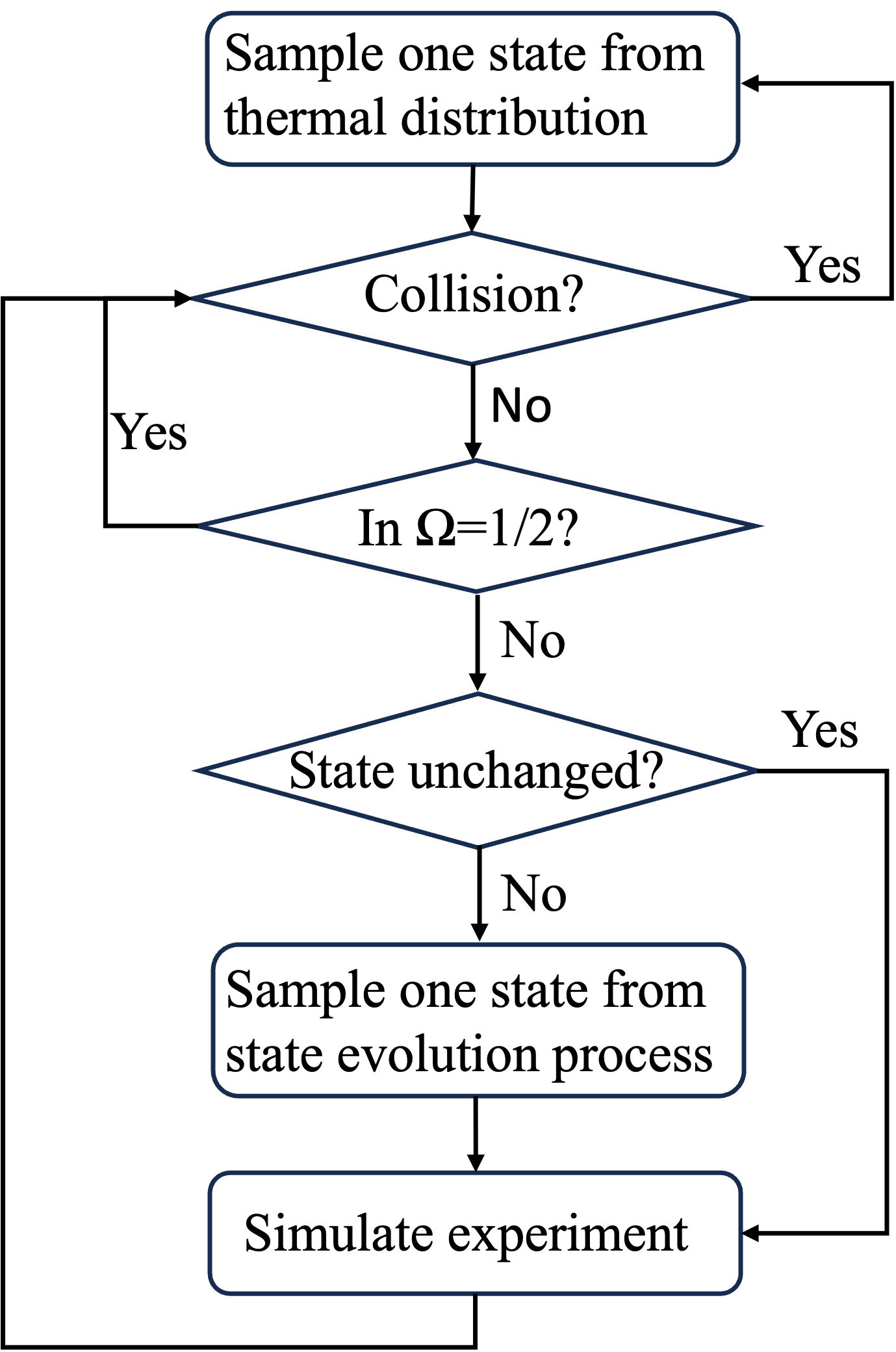}
    \end{minipage}\hfill
    \begin{minipage}[b]{0.542\linewidth}
        \includegraphics[width=\linewidth]{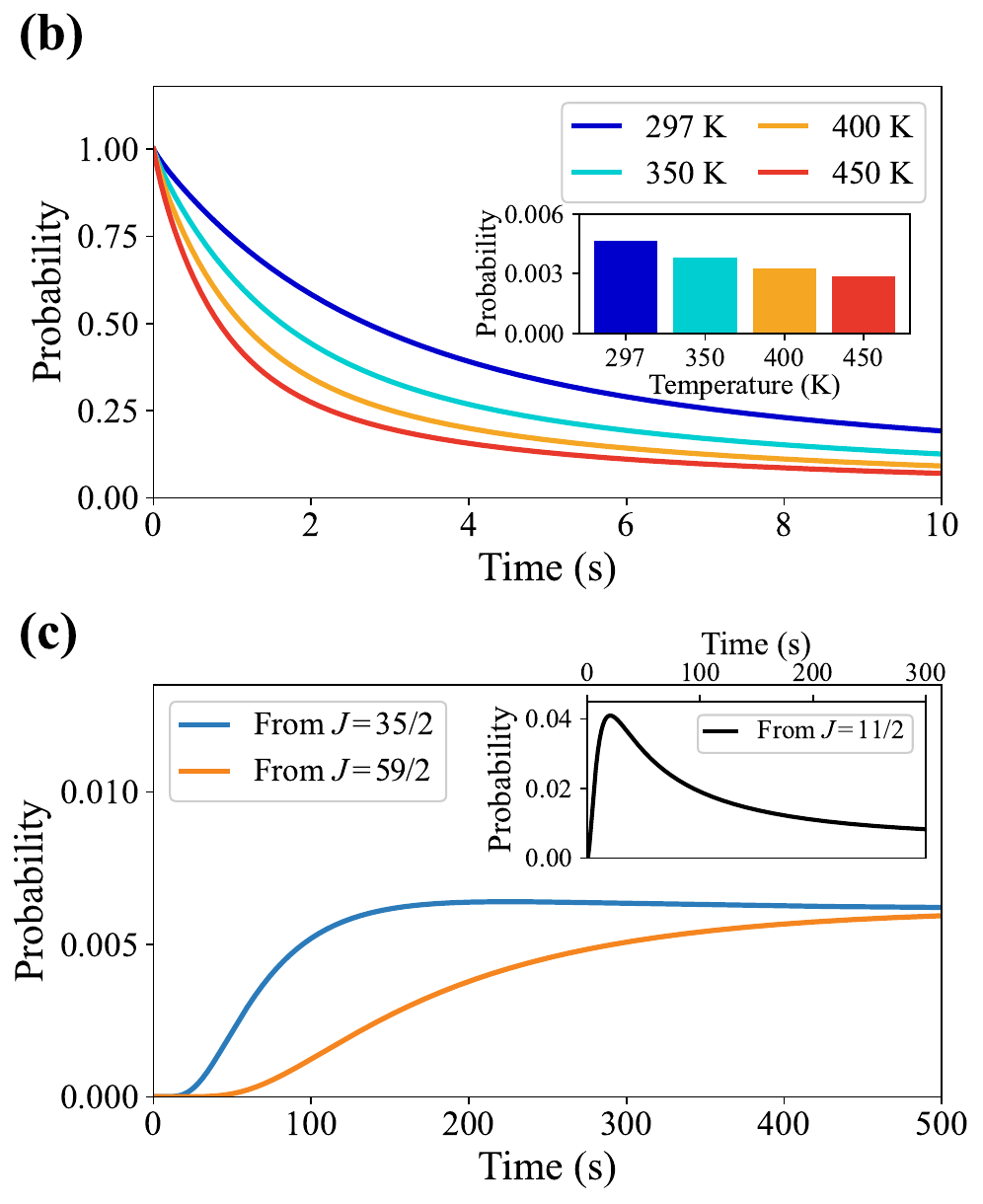}
    \end{minipage}
    \caption{Rotational-state model of CaO$^+$. (a) Flow chart of the Monte-Carlo simulation program that is used to track the rotational states of single CaO$^+$. The ``Simulate experiment'' step models the dipole-phonon interaction during the ramp without changing the rotational state. (b) Probability of finding the molecule in the ground rotational state at a given delay after it was prepared there, under different BBR temperatures; population that leaves the level and returns to it is included, so the $1/e$ time of $4.4$~s exceeds the $2.5$~s mean time to first leave. (Inset) Thermalized population of ground state under different BBR temperatures, normalized over both fine-structure manifolds. (c) Population evolution of ground rotational state with initial population at J=35/2 (most probable state at room temperature), J=59/2 and J=11/2 (Inset).}
    \label{fig:simulaitonlifetime}
    \label{fig:flowchart}
\end{figure*}

BBR continuously redistributes the vibrational and rotational populations, while transitions between the two fine-structure manifolds are electric-dipole-forbidden. The population dynamics follow the rate equation
\begin{widetext}
\begin{equation}
\begin{split}
\frac{dN_{v,j}}{dt} = &\sum_{E(v',j')>E(v,j)}^{v_{\text{max}},j_{\text{max}}} A_{(v',j')\rightarrow(v,j)} N_{v',j'} - \sum^{E(v',j')<E(v,j)}_{v_{0},j_{0}} A_{(v,j)\rightarrow(v',j')} N_{v,j} \\
&+ \sum_{v'=0,j'=0,(v',j')\neq (v,j)}^{v_{\text{max}},j_{\text{max}}}  \rho_{\text{BBR}}(|E_{v',j'}-E_{v,j}|, T) B_{(v',j')\rightarrow (v,j)}(N_{v',j'} - \frac{g_{j'}}{g_{j}}N_{v,j}),
\end{split}
\end{equation}
\end{widetext}
with $A$ and $B$ the Einstein coefficients and $\rho_{\text{BBR}}$ the Planck energy density. Because vibrational transitions dominate the BBR-driven dynamics, we evolve $70$ rotational states in each of the ground and first vibrational manifolds of $\Omega=3/2$. The ground rotational state re-thermalizes from the most-populated $J=35/2$ level in $\sim200$\,s (Fig.~\ref{fig:simulaitonlifetime}c); we adopt $300$\,s as the thermalization time of CaO$^+$.

In our model, transfer between the two fine-structure manifolds (\(\Omega=1/2\) and \(\Omega=3/2\)) occurs only through collisions with background gas, as radiative transitions between them are electric-dipole-forbidden. The total (Langevin) collision rate is set by the position-exchange rate of the two ions, 0.002\,s\(^{-1}\), from which a background pressure of 5\(\times 10^{-11}\)\,Torr can be inferred assuming Langevin collisions\cite{OKA2021111482}, consistent with our expectations. 
As a simplifying assumption, we take the internal-state reset rate to equal the estimated Langevin collision rate and draw the post-collision state from the Boltzmann distribution.
We use a Monte Carlo method to track the rotational state for each experiment. The flow chart of the simulation is shown in Fig.~\ref{fig:flowchart}a.
However, transfer between the two fine-structure manifolds requires inelastic, state-changing collisions, whose efficiency is not known for our system.
Rotational quenching has been estimated to occur only once per approximately 10–100 collisions in cold He buffer-gas systems\cite{GonzalezSanchez2018OHplus}. But this estimate does not directly determine the transfer rate across the 130~cm$^{-1}$ spin–orbit gap in CaO$^+$ under our experimental conditions.

\section{Experimental data and analysis methods}\label{app:data}
\subsection{Tail-based analysis}\label{app:tail}
To extract the ground-state population and per-event shelving probability (Fig.~\ref{fig:newcompare}a) we partition each dataset into non-overlapping $20$-experiment windows, so that no shot is shared between windows (the sliding-window binning used for the noise-floor bands reuses each shot in several bins). We nonetheless treat the window counts as independent, although a signal of the measured $1.9$~s duration covers about three windows. The CaO$^+$ count distribution is modeled as a mixture of an inactive, noise-only population that clicks with probability $p_n$ and the two DPQL-active $|m_J|$ manifolds. Angular momentum fixes the couplings of the two manifolds in a $3{:}1$ ratio, their populations are taken equal, and the steady-state simulation gives both per-event probabilities as a function of one effective coupling scale. The fit therefore varies only the total active population, the effective dipole and the noise floor $p_n$.
We fit by maximizing the Poisson likelihood of the binned counts over the full range of click numbers,
\begin{equation}
-\ln\mathcal{L}=\sum_k\left[\lambda_k-c_k\ln\lambda_k\right],\qquad \lambda_k=N_{\mathrm{win}}\,P_{\mathrm{model}}(k),
\end{equation}
where $c_k$ is the observed number of windows with $k$ events. The fit yields a total ground-rotational-state population of $0.60^{+0.14}_{-0.11}\%$, $p_s=0.17$ for $|m_J|=3/2$ and $0.02$ for $|m_J|=1/2$, corresponding to $d_{\text{PDM}}\approx2.59$~Debye.

The dipole-phonon signal is identified through the tail of the shelving-event distribution, which the noise floor produces only rarely. Here we focus on windows containing $\geq4$ events. For experiments with noise floor $p_n$ and $N_{\mathrm{win}}$ windows, the expected number of noise tail events is $\mu_n=N_{\mathrm{win}}(1-f)\,P(k\!\geq\!4\,|\,\mathrm{Binom}(20,p_n))$, where $f$ is the fraction of windows containing a molecule in the DPQL-active state, with Poisson fluctuation $\sqrt{\mu_n}$, defining a $3\sigma$ detection threshold $\mu_n+3\sqrt{\mu_n}$ in the Gaussian approximation. A single $p_n$ per dataset is not adequate here. The measured rate varies between runs by up to a factor of two, and $P(k\!\geq\!4\,|\,\mathrm{Binom}(20,p))$ rises as $p^{4}$ near our noise floor. So its average over runs exceeds its value at the average rate. As an example, a single $p_n$ underestimates $\mu_n$ by $19\%$ to $49\%$, most severely for the datasets whose background drifts most. We therefore measure $p_n$ run by run and sum $\mu_n$ over the runs of each dataset, with one run a contiguous block of $10^4$ experiments. The spread of $p_n$ between blocks also includes the scatter of a rate estimated from $N_b$ experiments, which has variance $p_n(1-p_n)/N_b$ even at a fixed rate. We remove this estimation noise before evaluating $\mu_n$, using the law of total variance. The resulting expectations are 32, 54, 41 and 31 windows for the middle-range, control-molecule, low- and high-range datasets, against 119, 60, 49 and 37 observed. The middle range is a detection at $11.7\sigma$, from the Poisson probability of the observed count under $\mu_n$. All three controls lie within $1.2\sigma$ of their expectations. The classification of all four datasets is unchanged for block lengths between $2\times10^3$ and $5\times10^4$ experiments. The threshold $k\geq4$ is chosen because lower thresholds are contaminated by noise, while the expected noise count at $k\geq4$ remains large enough ($\sim\!30$) that the threshold does not fall in the few-count regime.

\subsection{Hidden Markov model}
\label{app:HMMnew}
The posterior probability that an identified signal arises from the dipole-phonon interaction, given the prior probabilities, is computed via Bayes' theorem \cite{Lista_2016}:
\begin{equation}
\label{eq:ch7:BayesTheorem}
P(\Omega | x)=\frac{P(x | \Omega) P(\Omega)}{P(x)}=\frac{P(x | \Omega) P(\Omega)}{P(x | \Omega) P(\Omega)+P(x | J) P(J)}.
\end{equation}
Here $x$ is the observed record of Ca$^+$ fluorescence outcomes, $\Omega$ denotes the hypothesis that the molecule occupies the DPQL-active $|\Omega=3/2,J=3/2\rangle$ level, and $J$ the complementary hypothesis that it occupies any other rotational level. $P(x|\Omega)$ and $P(x|J)$ are the likelihoods of the record under the two hypotheses. $P(\Omega)$ and $P(J)=1-P(\Omega)$ are their prior probabilities, and $P(x)$ is the total probability of the record. The model assigns each experiment the posterior probability $P(\Omega|x)$. The hypothesis $\Omega$ covers the $J=3/2$ level as a whole, since the chain does not resolve $m_J$, so a single emission probability is attached to it, the measured $p_s=0.17$.
We evaluate it with a hidden Markov model (HMM) implemented in the hmmlearn Python package \cite{hmmlearn}, which predicts the probability that a measured observable state (Ca$^+$ fluorescence) was produced by a certain hidden state (CaO$^+$ rotation).
Its transition probabilities are the blackbody branching rates at $297$~K of Appendix~\ref{app:rotational}, which connect $70$ rotational levels in each of the $v=0$ and $v=1$ manifolds of $\Omega=3/2$ together with the $\Omega=1/2$ manifold.
We group these $212$ levels into $12$ hidden blocks, keeping separate every level that can be reached from $J=3/2$ in one step or return to it in one step, so that the fast vibrational return channel through $(v=1,J=3/2)$ is retained.
The model has no free parameters: the transition rates are fixed by the rate matrix, and the two emission probabilities are the measured per-event shelving probability $p_s=0.17$ and the noise floor $p_n=0.016$--$0.021$ of each dataset.
We read the signals from the posterior probability of each experiment rather than from the single most likely sequence of hidden states, which places $99.9\%$ of the experiments in the most weakly coupled block.

The dual-threshold criterion is calibrated on simulated data, in which the true state of the molecule is known.
It identifies no candidates in $33$~h of simulated background without signal. In simulations with signal, $99.8\%$ of the selected candidates overlap a true signal.
The criterion recovers $78\%$ of the simulated signals, so the signal counts quoted in the main text are lower bounds.
It also merges signals that are separated by a short return to the dark state, so that $707$ detected candidates cover $1050$ true signals.
The simulation result is decoded by the same HMM under the same dual-threshold criterion, and the durations it returns are compared directly with the measured ones (Fig.\,\ref{fig:hmmsummary}a).
Their $1/e$ times are $4920$~ms for the model and $1867^{+503}_{-337}$~ms for the data, a ratio of $2.6\pm0.6$.
With only $38$ events the crossing is set by the few durations nearest $1/e$, and the quoted interval spans the $22$nd to the $28$th longest.
The HMM can only detect candidate events with durations above a certain threshold.
As a result, the $4920$~ms here is the $1/e$ time of the candidate-event durations and is a different quantity from the $4.4$~s $1/e$ time of the $J=3/2$ occupancy in Fig.\,\ref{fig:simulaitonlifetime}b, which is the probability that the molecule is in the level at a given delay.

Signal rates are compared with an exact binomial test against the two off-resonant ranges. The common rate is removed by conditioning on the total. The number falling in the middle range is then distributed as $\mathrm{Binom}(N,\,T_{\mathrm{mid}}/T)$ with $T$ the total time. Of $42$ signals, $38$ fall in the middle range against the $14$ expected from its $13.1$~h of $38.1$~h. The rate ratio of middle to combination of low and high is $18$, with $p=5\times10^{-14}$.

\subsection{Joint frequency--coupling constraint}\label{app:joint}
How strongly the high- and low-range nulls constrain the model is set by the sensitivity of the tail test. An active-state signal $p_s$ would have produced an excess of $N_{\mathrm{win}}f\,P(k\!\geq\!4\,|\,\mathrm{Binom}(20,p_s))$ windows over the noise expectation $\mu_n$ of Appendix~\ref{app:tail}, which in units of the noise fluctuation is
\begin{equation}
Z_{\mathrm{exp}}(p_s)=\frac{N_{\mathrm{win}}\,f\,P(k\!\geq\!4\,|\,\mathrm{Binom}(20,p_s))}{\sqrt{\mu_n}},
\label{eq:zexp}
\end{equation}
with $f=0.3\%$ the fraction of windows containing a molecule in the DPQL-active state. This is a sensitivity rather than the significance of an observed excess.
The middle-range signal and the high- and low-range nulls are combined into a single goodness-of-fit over the effective dipole-phonon coupling $d_{\text{PDM}}$ and the doublet frequency $\omega_{\text{mol}}$ (Fig.~\ref{fig:rangebound}b). For each $(d_{\text{PDM}},\omega_{\text{mol}})$ the full simulation predicts the steady-state per-event shelving probability $p_{s,i}$ of each range, and we form
\begin{equation}
\chi^2=\left(\frac{p_{s,\mathrm{middle}}-0.17}{0.01}\right)^2+\sum_{\mathrm{high,\,low}}\!Z_{\mathrm{exp},i}(p_{s,i})^2,
\end{equation}
where $Z_{\mathrm{exp},i}$ is Eq.~\ref{eq:zexp} evaluated with the window count and noise expectation of range $i$. The first term enforces the measured shelving probability (two-sided, with the $0.01$ uncertainty from the fit above). The others penalize signals from the high and low ranges above their noise (one-sided). The minimum defines the best fit, and Fig.~\ref{fig:rangebound}b shows $\Delta\chi^2$ as a colour scale. We draw no confidence contour on it, since the coverage of a given level is not calibrated. The interval on $d_{\text{PDM}}$ is the $\Delta\chi^2\leq1$ profile, $\pm0.07$~Debye. The two nulls confine $\omega_{\text{mol}}/2\pi$ to $453.2$--$454.5$~kHz, for which we quote $2\pi\times454\pm1$~kHz. The middle-range data constrain $d_{\text{PDM}}$, while the high- and low-range data bound $\omega_{\text{mol}}$ from above and below. Together they isolate $\omega_{\text{mol}}=2\pi\times454\pm1$~kHz and $d_{\text{PDM}}\approx2.56$~Debye.

\bibliography{DPQL}

@article{kortunov2021proton,
  title={Proton--electron mass ratio by high-resolution optical spectroscopy of ion ensembles in the resolved-carrier regime},
  author={Kortunov, IV and Alighanbari, S and Hansen, MG and Giri, GS and Korobov, VI and Schiller, S},
  journal={Nature physics},
  volume={17},
  number={5},
  pages={569--573},
  year={2021},
  publisher={Nature Publishing Group}
}

@article{Qi2023,
  title = {Adiabatically controlled motional states of a CaO$^{+}$ and Ca$^{+}$ trapped-ion chain cooled to the ground state},
  author = {Qi, Lu and Reed, Evan C. and Brown, Kenneth R.},
  journal = {Phys. Rev. A},
  volume = {108},
  issue = {1},
  pages = {013108},
  numpages = {9},
  year = {2023},
  month = {Jul},
  publisher = {American Physical Society},
  doi = {10.1103/PhysRevA.108.013108},
  url = {https://link.aps.org/doi/10.1103/PhysRevA.108.013108}
}

@article{Chou2017, title={Preparation and coherent manipulation of pure quantum states of a single molecular ion}, volume={545}, ISSN={1476-4687}, DOI={10.1038/nature22338}, number={7653}, journal={Nature}, author={Chou, Chin-wen and Kurz, Christoph and Hume, David B. and Plessow, Philipp N. and Leibrandt, David R. and Leibfried, Dietrich}, year={2017}, month=may, pages={203–207} }

@article{Wolf2016, title={Non-destructive state detection for quantum logic spectroscopy of molecular ions}, volume={530}, ISSN={1476-4687}, DOI={10.1038/nature16513}, number={7591}, journal={Nature}, author={Wolf, Fabian and Wan, Yong and Heip, Jan C. and Gebert, Florian and Shi, Chunyan and Schmidt, Piet O.}, year={2016}, month=feb, pages={457–460} }

@article{Campbell_Hudson_2020,
  title = {Dipole-Phonon Quantum Logic with Trapped Polar Molecular Ions},
  author = {Campbell, Wesley C. and Hudson, Eric R.},
  journal = {Phys. Rev. Lett.},
  volume = {125},
  issue = {12},
  pages = {120501},
  numpages = {6},
  year = {2020},
  month = {Sep},
  publisher = {American Physical Society},
  doi = {10.1103/PhysRevLett.125.120501},
  url = {https://link.aps.org/doi/10.1103/PhysRevLett.125.120501}
}

@article{Schmidt_QLS_2005, title={Spectroscopy Using Quantum Logic}, volume={309}, DOI={10.1126/science.1114375}, abstractNote={}, number={5735}, journal={Science}, author={Schmidt, P. O. and Rosenband, T. and Langer, C. and Itano, W. M. and Bergquist, J. C. and Wineland, D. J.}, year={2005}, pages={749–752} }

@article{shuman2010laser,
  title={Laser cooling of a diatomic molecule},
  author={Shuman, Edward S and Barry, John F and DeMille, David},
  journal={Nature},
  volume={467},
  number={7317},
  pages={820--823},
  year={2010},
  publisher={Nature Publishing Group}
}

@article{
doi:10.1126/science.aaz9837,
author = {Mudit Sinhal  and Ziv Meir  and Kaveh Najafian  and Gregor Hegi  and Stefan Willitsch },
title = {Quantum-nondemolition state detection and spectroscopy of single trapped molecules},
journal = {Science},
volume = {367},
number = {6483},
pages = {1213-1218},
year = {2020},
doi = {10.1126/science.aaz9837},
URL = {https://www.science.org/doi/abs/10.1126/science.aaz9837},
eprint = {https://www.science.org/doi/pdf/10.1126/science.aaz9837}}

@article{mills2020dipole,
  title={Dipole--phonon quantum logic with alkaline-earth monoxide and monosulfide cations},
  author={Mills, Michael and Wu, Hao and Reed, Evan C and Qi, Lu and Brown, Kenneth R and Schneider, Christian and Heaven, Michael C and Campbell, Wesley C and Hudson, Eric R},
  journal={Physical Chemistry Chemical Physics},
  volume={22},
  number={43},
  pages={24964--24973},
  year={2020},
  publisher={Royal Society of Chemistry}
}

@article{Vogelius_2006,
doi = {10.1088/0953-4075/39/19/S31},
url = {https://dx.doi.org/10.1088/0953-4075/39/19/S31},
year = {2006},
month = {sep},
publisher = {},
volume = {39},
number = {19},
pages = {S1259},
author = {I S Vogelius and L B Madsen and M Drewsen},
title = {Probabilistic state preparation of a single molecular ion by projection measurement},
journal = {Journal of Physics B: Atomic, Molecular and Optical Physics}
}

@article{Meekhof_Wineland_1996,
  title = {Generation of Nonclassical Motional States of a Trapped Atom},
  author = {Meekhof, D. M. and Monroe, C. and King, B. E. and Itano, W. M. and Wineland, D. J.},
  journal = {Phys. Rev. Lett.},
  volume = {76},
  issue = {11},
  pages = {1796--1799},
  numpages = {0},
  year = {1996},
  month = {Mar},
  publisher = {American Physical Society},
  doi = {10.1103/PhysRevLett.76.1796},
  url = {https://link.aps.org/doi/10.1103/PhysRevLett.76.1796}
}

@article{Schmid_2019, title={Quantum-state-specific reaction rate measurements for the photo-induced reaction Ca+ + O2 → CaO+ + O}, volume={117}, ISSN={0026-8976}, DOI={10.1080/00268976.2019.1622811}, abstractNote={}, number={21}, journal={Molecular Physics}, publisher={Taylor & Francis}, author={Schmid, Philipp C. and Miller, Mikhail I. and Greenberg, James and Nguyen, Thanh L. and Stanton, John F. and Lewandowski, H. J.}, year={2019}, month=nov, pages={3036–3042} }

@article{acme2018improved,
  title={Improved limit on the electric dipole moment of the electron},
  author={V. Andreev and et. al.},
  collaboration={ACME Collaboration},
  journal={Nature},
  volume={562},
  number={7727},
  pages={355--360},
  year={2018},
  publisher={Nature Publishing Group UK London}
}

@article{
doi:10.1126/science.adg4084,
author = {Tanya S. Roussy  and Luke Caldwell  and Trevor Wright  and William B. Cairncross  and Yuval Shagam  and Kia Boon Ng  and Noah Schlossberger  and Sun Yool Park  and Anzhou Wang  and Jun Ye  and Eric A. Cornell },
title = {An improved bound on the electron's electric dipole moment},
journal = {Science},
volume = {381},
number = {6653},
pages = {46-50},
year = {2023},
doi = {10.1126/science.adg4084},
URL = {https://www.science.org/doi/abs/10.1126/science.adg4084},
eprint = {https://www.science.org/doi/pdf/10.1126/science.adg4084}}

@article{karman2024ultracold,
  title={Ultracold chemistry as a testbed for few-body physics},
  author={Karman, Tijs and Tomza, Micha{\l} and P{\'e}rez-R{\'\i}os, Jes{\'u}s},
  journal={Nature Physics},
  pages={1--8},
  year={2024},
  publisher={Nature Publishing Group UK London}
}

@article{chen2023field,
  title={Field-linked resonances of polar molecules},
  author={Chen, Xing-Yan and Schindewolf, Andreas and Eppelt, Sebastian and Bause, Roman and Duda, Marcel and Biswas, Shrestha and Karman, Tijs and Hilker, Timon and Bloch, Immanuel and Luo, Xin-Yu},
  journal={Nature},
  volume={614},
  number={7946},
  pages={59--63},
  year={2023},
  publisher={Nature Publishing Group UK London}
}

@article{
doi:10.1126/science.adf8999,
author = {Yicheng Bao  and Scarlett S. Yu  and LoÃ¯c Anderegg  and Eunmi Chae  and Wolfgang Ketterle  and Kang-Kuen Ni  and John M. Doyle },
title = {Dipolar spin-exchange and entanglement between molecules in an optical tweezer array},
journal = {Science},
volume = {382},
number = {6675},
pages = {1138-1143},
year = {2023},
doi = {10.1126/science.adf8999},
URL = {https://www.science.org/doi/abs/10.1126/science.adf8999},
eprint = {https://www.science.org/doi/pdf/10.1126/science.adf8999}}

@article{li2023tunable,
  title={Tunable itinerant spin dynamics with polar molecules},
  author={Li, Jun-Ru and Matsuda, Kyle and Miller, Calder and Carroll, Annette N and Tobias, William G and Higgins, Jacob S and Ye, Jun},
  journal={Nature},
  volume={614},
  number={7946},
  pages={70--74},
  year={2023},
  publisher={Nature Publishing Group UK London}
}

@article{
doi:10.1126/science.adl6570,
author = {Yi-Xiang Liu  and Lingbang Zhu  and Jeshurun Luke  and J. J. Arfor Houwman  and Mark C. Babin  and Ming-Guang Hu  and Kang-Kuen Ni },
title = {Quantum interference in atom-exchange reactions},
journal = {Science},
volume = {384},
number = {6700},
pages = {1117-1121},
year = {2024},
doi = {10.1126/science.adl6570},
URL = {https://www.science.org/doi/abs/10.1126/science.adl6570},
eprint = {https://www.science.org/doi/pdf/10.1126/science.adl6570}}

@article{
doi:10.1126/science.adf4272,
author = {Connor M. Holland  and Yukai Lu  and Lawrence W. Cheuk },
title = {On-demand entanglement of molecules in a reconfigurable optical tweezer array},
journal = {Science},
volume = {382},
number = {6675},
pages = {1143-1147},
year = {2023},
doi = {10.1126/science.adf4272},
URL = {https://www.science.org/doi/abs/10.1126/science.adf4272},
eprint = {https://www.science.org/doi/pdf/10.1126/science.adf4272}}

@article{lin2020quantum,
  title={Quantum entanglement between an atom and a molecule},
  author={Lin, Yiheng and Leibrandt, David R and Leibfried, Dietrich and Chou, Chin-wen},
  journal={Nature},
  volume={581},
  number={7808},
  pages={273--277},
  year={2020},
  publisher={Nature Publishing Group UK London}
}

@article{
doi:10.1126/science.aan4701,
author = {Prateek Puri  and Michael Mills  and Christian Schneider  and Ionel Simbotin  and John A. Montgomery  and Robin CÃŽtÃ©  and Arthur G. Suits  and Eric R. Hudson },
title = {Synthesis of mixed hypermetallic oxide BaOCa<sup>+</sup> from laser-cooled reagents in an atom-ion hybrid trap},
journal = {Science},
volume = {357},
number = {6358},
pages = {1370-1375},
year = {2017},
doi = {10.1126/science.aan4701},
URL = {https://www.science.org/doi/abs/10.1126/science.aan4701},
eprint = {https://www.science.org/doi/pdf/10.1126/science.aan4701}}

@article{
doi:10.1126/science.aba3628,
author = {C. W. Chou  and A. L. Collopy  and C. Kurz  and Y. Lin  and M. E. Harding  and P. N. Plessow  and T. Fortier  and S. Diddams  and D. Leibfried  and D. R. Leibrandt },
title = {Frequency-comb spectroscopy on pure quantum states of a single molecular ion},
journal = {Science},
volume = {367},
number = {6485},
pages = {1458-1461},
year = {2020},
doi = {10.1126/science.aba3628},
URL = {https://www.science.org/doi/abs/10.1126/science.aba3628},
eprint = {https://www.science.org/doi/pdf/10.1126/science.aba3628}}

@article{holzapfel2024quantum,
  title={Quantum control of a single $\mathrm{H}_2^+$
      molecular ion},
  author={Holzapfel, David and Schmid, Fabian and Schwegler, Nick and Stadler, Oliver and Stadler, Martin and Ferk, Alexander and Home, Jonathan P and Kienzler, Daniel},
  journal={arXiv preprint arXiv:2409.06495},
  year={2024}
}

@article{staanum2010rotational,
  title={Rotational laser cooling of vibrationally and translationally cold molecular ions},
  author={Staanum, Peter F and H{\o}jbjerre, Klaus and Skyt, Peter S and Hansen, Anders K and Drewsen, Michael},
  journal={Nature Physics},
  volume={6},
  number={4},
  pages={271--274},
  year={2010},
  publisher={Nature Publishing Group UK London}
}

@article{germann2014observation,
  title={Observation of electric-dipole-forbidden infrared transitions in cold molecular ions},
  author={Germann, Matthias and Tong, Xin and Willitsch, Stefan},
  journal={Nature Physics},
  volume={10},
  number={11},
  pages={820--824},
  year={2014},
  publisher={Nature Publishing Group UK London}
}

@article{calvin2018rovibronic,
  title={Rovibronic spectroscopy of sympathetically cooled 40CaH+},
  author={Calvin, Aaron T and Janardan, Smitha and Condoluci, John and Rugango, Rene and Pretzsch, Eric and Shu, Gang and Brown, Kenneth R},
  journal={The Journal of Physical Chemistry A},
  volume={122},
  number={12},
  pages={3177--3181},
  year={2018},
  publisher={ACS Publications}
}

@article{VANGUNDY201817,
title = {Spectroscopy of the low-lying states of CaO+},
journal = {Journal of Molecular Spectroscopy},
volume = {344},
pages = {17-20},
year = {2018},
issn = {0022-2852},
doi = {https://doi.org/10.1016/j.jms.2017.10.001},
url = {https://www.sciencedirect.com/science/article/pii/S0022285217303776},
author = {Robert A. VanGundy and Joshua H. Bartlett and Michael C. Heaven}
}

@article{
doi:10.1126/science.ado1001,
author = {Yu Liu  and Julian Schmidt  and Zhimin Liu  and David R. Leibrandt  and Dietrich Leibfried  and Chin-wen Chou },
title = {Quantum state tracking and control of a single molecular ion in a thermal environment},
journal = {Science},
volume = {385},
number = {6710},
pages = {790-795},
year = {2024},
doi = {10.1126/science.ado1001},
URL = {https://www.science.org/doi/abs/10.1126/science.ado1001},
eprint = {https://www.science.org/doi/pdf/10.1126/science.ado1001}}

@article{OKA2021111482,
title = {Forbidden rotational transitions and astrophysics},
journal = {Journal of Molecular Spectroscopy},
volume = {379},
pages = {111482},
year = {2021},
issn = {0022-2852},
doi = {https://doi.org/10.1016/j.jms.2021.111482},
url = {https://www.sciencedirect.com/science/article/pii/S0022285221000667},
author = {Takeshi Oka}
}

@article{10.1116/1.570570,
    author = {Fulford, John Edward and Nhu‐Hoa, Duong and Hughes, Richard James and March, Raymond Evans and Bonner, Ronald Francis and Wong, Gordon John},
    title = "{Radio‐frequency mass selective excitation and resonant ejection of ions in a three‐dimensional quadrupole ion trap}",
    journal = {Journal of Vacuum Science and Technology},
    volume = {17},
    number = {4},
    pages = {829-835},
    year = {1980},
    month = {07},
    issn = {0022-5355},
    doi = {10.1116/1.570570},
    url = {https://doi.org/10.1116/1.570570},
    eprint = {https://pubs.aip.org/avs/jvst/article-pdf/17/4/829/11532045/829\_1\_online.pdf},
}

@online{hmmlearn,
  author        = {Weiss, Ron and Du, Shiqiao and Grobler, Jaques and Cournapeau, David and Pedregosa, Fabian and Varoquaux, Gael and Mueller, Andreas and Thirion, Bertrand and Nouri, Daniel and Louppe, Gilles and Vanderplas, Jake and Benediktsson, John and Buitinck, Lars and Korobov, Mikhail and McGibbon, Robert and Lattarini, Stefano and Niculae, V. and Gramfort, Alexandre and Lebedev, Sergei Huppenkothen, Daniela and Farrow, Christopher and Yanenko, Alexandr and Lee,Antony and Danielson, Matthew and Rockhill, Alex},
  title         = {hmmlearn},
  year          = "2024",
  note          = {{https://github.com/hmmlearn/hmmlearn}}
}

@book{Lista_2016,
  title={Statistical Methods for Data Analysis in Particle Physics},
  author={Lista, Luca},
  isbn={978-3-319-20175-7},
  series={Lecture Notes in Physics},
  url={https://link.springer.com/10.1007/978-3-319-20176-4},
  year={2016},
  publisher={Springer International Publishing}
}

@article{7ypf-91jr,
  title = {High-Fidelity Quantum State Control of a Polar Molecular Ion in a Cryogenic Environment},
  author = {Chaffee, Dalton and Margulis, Baruch and Sheffield, April and Schmidt, Julian and Reisenfeld, April and Leibrandt, David R. and Leibfried, Dietrich and Chou, Chin-Wen},
  journal = {Phys. Rev. Lett.},
  volume = {135},
  issue = {24},
  pages = {240801},
  numpages = {8},
  year = {2025},
  month = {Dec},
  publisher = {American Physical Society},
  doi = {10.1103/7ypf-91jr},
  url = {https://link.aps.org/doi/10.1103/7ypf-91jr}
}

@article{doi:10.1021/acs.jpca.5c00229,
author = {Patel, Swapnil and Brown, Kenneth R.},
title = {Precise Determination of Excited State Rotational Constants and Black-Body Thermometry in Coulomb Crystals of Ca+ and CaH+},
journal = {The Journal of Physical Chemistry A},
volume = {129},
number = {16},
pages = {3624-3629},
year = {2025},
doi = {10.1021/acs.jpca.5c00229},
    note ={PMID: 40237448},

URL = { 
    
        https://doi.org/10.1021/acs.jpca.5c00229
    
    

},
eprint = { 
    
        https://doi.org/10.1021/acs.jpca.5c00229
    
    

}

}

@article{PhysRevA.89.062332,
  title = {Two-mode coupling in a single-ion oscillator via parametric resonance},
  author = {Gorman, Dylan J and Schindler, Philipp and Selvarajan, Sankaranarayanan and Daniilidis, Nikos and H\"affner, Hartmut},
  journal = {Phys. Rev. A},
  volume = {89},
  issue = {6},
  pages = {062332},
  numpages = {7},
  year = {2014},
  month = {Jun},
  publisher = {American Physical Society},
  doi = {10.1103/PhysRevA.89.062332},
  url = {https://link.aps.org/doi/10.1103/PhysRevA.89.062332}
}

@article{PhysRevLett.119.193602,
  title = {Cross-Kerr Nonlinearity for Phonon Counting},
  author = {Ding, Shiqian and Maslennikov, Gleb and Habl\"utzel, Roland and Matsukevich, Dzmitry},
  journal = {Phys. Rev. Lett.},
  volume = {119},
  issue = {19},
  pages = {193602},
  numpages = {5},
  year = {2017},
  month = {Nov},
  publisher = {American Physical Society},
  doi = {10.1103/PhysRevLett.119.193602},
  url = {https://link.aps.org/doi/10.1103/PhysRevLett.119.193602}
}

@misc{bohn2009electricdipolesultralowtemperatures,
      title={Electric dipoles at ultralow temperatures},
      author={John L. Bohn},
      year={2009},
      eprint={0901.0276},
      archivePrefix={arXiv},
      primaryClass={physics.atom-ph},
      url={https://arxiv.org/abs/0901.0276},
}

@misc{Qi2024evidence,
  title = {Experimental evidence for dipole-phonon quantum logic in a trapped calcium monoxide and calcium ion chain},
  author = {Qi, Lu and Reed, Evan C. and Yu, Boyan and Brown, Kenneth R.},
  year = {2024},
  eprint = {2411.07137},
  archivePrefix = {arXiv},
  primaryClass = {quant-ph},
  url = {https://arxiv.org/abs/2411.07137}
}

@article{GonzalezSanchez2018OHplus,
  title = {Modeling Quantum Kinetics in Ion Traps: State-changing Collisions for OH$^+$($^3\Sigma^-$) Ions with He as a Buffer Gas},
  author = {Gonz{\'a}lez-S{\'a}nchez, L. and Wester, R. and Gianturco, F. A.},
  journal = {ChemPhysChem},
  volume = {19},
  issue = {15},
  pages = {1866--1875},
  year = {2018},
  doi = {10.1002/cphc.201800119},
  url = {https://doi.org/10.1002/cphc.201800119}
}

@article{wu2026infrared,
  title={Infrared absorption spectroscopy of a single polyatomic molecular ion},
  author={Wu, Zhenlin and Duka, Tim and Isaza-Monsalve, Mariano and Kautzky, Miriam and {\v{S}}varc, Vojt{\v{e}}ch and Turci, Andrea and Nardi, Ren{\'e} and Gronowski, Marcin and Tomza, Micha{\l} and Furey, Brandon J. and Schindler, Philipp},
  journal={Nature},
  volume={657},
  number={8131},
  pages={364--368},
  year={2026},
  doi={10.1038/s41586-026-10915-8}
}

\end{document}